\documentclass[11pt,a4paper]{article}

\usepackage[utf8]{inputenc}
\usepackage[T1]{fontenc}
\usepackage[english]{babel}
\usepackage{lmodern}
\usepackage{microtype}
\usepackage{geometry}
\usepackage{booktabs}
\usepackage{tabularx}
\usepackage{array}
\usepackage{amsmath}
\usepackage{graphicx}
\graphicspath{{figures/}}
\usepackage{hyperref}
\usepackage{url}
\usepackage{parskip}

\hypersetup{
  colorlinks=true,
  linkcolor=blue,
  citecolor=blue,
  urlcolor=blue
}

\newcolumntype{Y}{>{\raggedright\arraybackslash}X}

\title{What makes a \textit{harness} a \textit{harness}: necessary and sufficient conditions for an agent harness}
\author{Sanderson Oliveira de Macedo\\
\textit{Federal Institute of Goiás}\\
\texttt{sanderson.macedo@ifg.edu.br}}
\date{\today}

\begin{document}

\maketitle

\begin{abstract}
The term \textit{agent harness} now circulates widely in software engineering with generative artificial intelligence. It names the layer that wraps a language model and turns it into a coding agent able to act on a repository. The usage is loose and polysemous. Sometimes the term denotes the whole product (Claude Code, Codex CLI); sometimes it denotes the evaluation scaffold that runs an agent against tasks (the SWE-bench harness); sometimes it gets conflated with an agent framework, an SDK, an IDE plugin, or an orchestrator. What is missing is a reference definition that works as an instrument, one that includes and excludes cases consistently. We build that definition through a conceptual analysis that combines works with persistent identifiers and primary grey-literature sources, such as official documentation, glossaries, and engineering reports. We reconstruct the genealogy of the term, from the horse's tack to the classic \textit{test harness}, to the machine-learning \textit{evaluation harness}, and finally to the \textit{agent harness}. We then propose a constitutive definition that states the necessary and sufficient conditions for a system to be an agent harness, we operationalize it as an inclusion and exclusion test, and we draw the boundary of the concept against an agent framework, an agent SDK, an IDE plugin, an eval harness, and an orchestrator. We apply the definition to six real harnesses (Claude Code, Codex CLI, Aider, Cline, OpenHands, and SWE-agent) and to deliberate edge cases; the test includes and excludes consistently. We close with a research agenda organized by design tension axes. The contribution is an operational definition of agent harness, with a shared vocabulary, able to guide engineering practice and the scientific comparison of agentic systems.
\end{abstract}

\noindent\textbf{Keywords:} agent harness; coding agent; software engineering with generative AI; agent-computer interface; context engineering; operational definition.

\section{Introduction}

Picture a draft horse. It has power to spare. What it lacks is direction: loose, it bolts; tied to nothing, it moves no load at all. What turns brute force into useful work is the tack, the set of straps and belts that ties the animal to the cart, transmits the pull, steers the course, and stops the bolt. The tack does not replace the horse, and it does not compete with it. It works on a different layer, the engineering that channels power safely. English fixed this image in the verb \textit{to harness}, and the image fits an object that has become central in recent software engineering: the layer that wraps a language model and turns it into an agent able to write code.

Large language models are no longer mere text generators. They have become the engine of systems that decide, call tools, and act on the environment \cite{Brown2020FewShot,Wei2022Emergent,Zhao2023SurveyLLM}. The decisive turn was to couple reasoning and action. Instruction alignment made models follow requests \cite{Ouyang2022InstructGPT}; chain-of-thought, tree-of-thought, and self-consistency improved deliberation \cite{Wei2022CoT,Yao2023ToT,Wang2023SelfConsistency}; patterns such as ReAct came to interleave thinking and acting in a loop; and external tools, from function calling to browsing, extended the model's reach beyond its context window \cite{Yao2023ReAct,Schick2023Toolformer,Qin2023ToolLearning}. On that base, coding agents were born. They take a task in natural language, read a repository, edit files, run commands, and verify their own work \cite{Yang2024SWEagent,Wang2024OpenHands,Zhang2024AutoCodeRover}. The literature already treats these agents as a research line of their own, with surveys that systematize their architectures, capabilities, and limitations \cite{Wang2024SurveyAutonomous,Xi2025RisePotential,Liu2024AgentsSE}.

In this emerging vocabulary, one word keeps surfacing to name the infrastructure that surrounds the model: \textit{harness}. The official Claude Code documentation defines the \textit{agentic harness} as the set of tools, context management, and execution environment that turns a model into a capable coding agent, and sums up the relationship in one sentence: Claude Code is the harness, and Claude is the model inside it.\footnote{Official Claude Code documentation, entry \textit{Agentic harness}: \url{https://code.claude.com/docs/en/glossary}.} The Hugging Face agent glossary records the same broad usage, naming products such as Claude Code, Codex, and Antigravity as harnesses, that is, everything that is not the model.\footnote{Hugging Face, \textit{Harness, Scaffold, and the AI Agent Terms Worth Getting Right}: \url{https://huggingface.co/blog/agent-glossary}.} The term travels without a passport. In classic machine learning, \textit{harness} denotes an evaluation suite; SWE-bench itself, today the reference evaluation for coding agents, calls its execution scaffold a harness \cite{Jimenez2024SWEbench}. In software engineering, \textit{test harness} predates generative AI. And in agentic usage, \textit{harness} sometimes means the whole product, sometimes the orchestration layer, sometimes gets conflated with an agent framework, an SDK, an IDE plugin, or an orchestrator. The polysemy is not a terminological detail. It blocks us from comparing systems, attributing design merit, and accumulating knowledge about what, after all, makes an agent trustworthy.

The practical consequence of this confusion has a name. An agent hits an obstacle and, instead of admitting it, reports a success that did not happen. The common reflex is to tweak the prompt. But language models, when cornered, tend to produce the answer that seems to satisfy the request, true or not \cite{Bubeck2023Sparks}, and politely instructing the model not to do so is the weakest of controls. The robust solution is engineering around the model: detect the divergence between what the agent claims and the real state, verify that state deterministically, and run the sensitive parts with ordinary code rather than trust the model's word. It is that layer, not the prompt, that separates a demo that impresses from a product that holds up. Giving it a precise name is the object of this article.

The concept is central in practice, yet the academic literature rarely defines it head-on. Surveys of software agents describe components (memory, tools, reasoning loop) without consolidating what constitutes the harness as a unit \cite{Hou2024LLM4SE,Liu2024AgentsSE,Masterman2024Landscape}; work on the agent-computer interface treats one facet, tools, and stops there \cite{Yang2024SWEagent}; context engineering, now a topic of its own, covers another facet, context management, and does not close the definition \cite{Mei2025ContextEng}. The reference definition that works as an instrument is missing: one that says, given a concrete system, whether or not it is an agent harness, and why.

This article delivers that definition. The central question is direct: what is the reference definition of \textit{agent harness}, and how do we make it an instrument that discriminates cases rather than a slogan? To answer it, we organize five contributions around five research questions. \textbf{RQ1 (genealogy):} where does the term \textit{harness} come from and how did its sense migrate to agentic usage? \textbf{RQ2 (constitutive definition):} what conditions are necessary and sufficient for a system to be an agent harness? \textbf{RQ3 (boundary):} how does the definition distinguish a harness from an agent framework, an SDK, an IDE plugin, an eval harness, and an orchestrator? \textbf{RQ4 (application):} how do real harnesses instantiate the constitutive conditions? \textbf{RQ5 (agenda):} which design axes remain open?

The contributions are, in order: a genealogy that reconstructs the term's sense migration (Section~\ref{sec:genealogia}); a constitutive definition operationalized as an inclusion and exclusion test (Section~\ref{sec:definicao}); a boundary delimitation against five neighboring concepts (Section~\ref{sec:fronteira}); the application of the definition to six real harnesses, showing that it classifies consistently (Section~\ref{sec:aplicacao}); and a research agenda organized by tension axes (Section~\ref{sec:agenda}). Section~\ref{sec:relacionados} positions the article against the literature.

Positioning. This article is deliberately definitional. It does not propose a new agent, does not measure benchmark performance, and does not recapitulate taxonomies of AI systems for software engineering that the literature already has. Its product is conceptual: a shared vocabulary and a reproducible test of membership in the concept. The wager is simple. Terminological clarity is a precondition for cumulative science in a field that, today, calls too many things a \textit{harness}. A note on method: this is a work of conceptual clarification, not a survey or a usage count; each source with a verified DOI enters as a formal citation, while product documentation and glossaries, where the term is actually used, enter as footnotes with a URL.

\section{Related Work}
\label{sec:relacionados}

The literature around the agent harness is vast, but it touches the concept sideways, by facets, and never closes it. Surveys of language agents describe architectures in terms of perception, memory, planning, and action, and recent studies shift the focus to evaluating agentic behavior \cite{Wang2024SurveyAutonomous,Xi2025RisePotential,Sumers2024Cognitive,Yehudai2025AgentEval,Chang2024SurveyEval}; they treat the agent as the object and push the infrastructure into the background. In software engineering, maps of model use across the lifecycle \cite{Hou2024LLM4SE,Zhang2023NLPSE,Liu2024AgentsSE}, agentic runtime systems \cite{Yang2024SWEagent,Wang2024OpenHands,Zhang2024AutoCodeRover,Xia2024Agentless,Chen2024CodeR,Arora2024MASAI}, multi-agent architectures \cite{Qian2024ChatDev,Hong2024MetaGPT,Wu2023AutoGen,Li2023CAMEL,Chen2023AgentVerse,Fourney2024MagenticOne,Gao2024AgentScope,Zhou2023AgentsFramework}, and studies of code action and collaboration \cite{Wang2024CodeAct,Dong2023SelfCollaboration,Park2023GenerativeAgents} each describe their own system, without the cross-cutting axis. Another front goes deep into isolated components, the agent-computer interface, context engineering, memory, tool use, and self-correction \cite{Mei2025ContextEng,Liu2024LostMiddle,Lewis2020RAG,Zhang2024MemorySurvey,Packer2023MemGPT,Wang2024WorkflowMemory,Qin2024ToolLLM,Li2023APIBank,Song2023RestGPT,Shen2023HuggingGPT,Madaan2023SelfRefine,Shinn2023Reflexion}. And the control front deals with guardrails, attacks, and benchmark evaluation \cite{Rebedea2023NeMo,Inan2023LlamaGuard,Barrett2023SecurityRisks,Liu2023PromptInjection,Debenedetti2024AgentDojo,Wei2023Jailbroken,Zou2023Adversarial,Jimenez2024SWEbench,Yang2024SWEbenchMM,Liu2024AgentBench,Zhou2024WebArena,Deng2023Mind2Web,Yao2024TauBench,Liu2023EvalPlus,Bai2022ConstitutionalAI}. Each piece is part of the harness; none, alone, is the harness. What is missing is a reference definition that unites these fronts into a concept with sharp boundaries and a membership test. That is the gap we occupy. This article does not compete with the system catalogs, which say what; it answers the \textit{what it is}, prior to any catalog. The systems reappear as empirical material in Section~\ref{sec:aplicacao}, and the evaluation sense of the term is separated from the agentic sense in Section~\ref{sec:fronteira}.

\section{Genealogy of the Term (RQ1)}
\label{sec:genealogia}

To define a term well, start with where it came from. The word \textit{harness} carries a largely stable metaphor across centuries and domains. Reconstruct that metaphor and you see why it was repurposed to name the infrastructure of agents. We answer RQ1 by tracing four stations: the etymological origin, the \textit{test harness} of software engineering, the \textit{evaluation harness} of machine learning, and the \textit{agent harness}.

\subsection*{From armor to tack}
In English, \textit{harness} comes from the Old French \textit{harneis} (also \textit{harnois}, twelfth century): equipment, armor, the set of war gear. Its origin is uncertain, possibly the Old Norse \textit{hernest}, the provisions of an army. Around 1300, the word enters English meaning fighting equipment and armor, including that of a knight \textit{in full harness}, in complete armor. By the early fourteenth century the non-military sense leaves the battlefield for the stable: \textit{harness} comes to name the set of straps and belts that connects a draught animal to a cart, the tack.\footnote{Dates and senses follow historical dictionaries of English: \textit{Online Etymology Dictionary}, entry \textit{harness} (\url{https://www.etymonline.com/word/harness}); \textit{Oxford English Dictionary}, entries \textit{harness, n.} and \textit{harness, v.}; and \textit{Merriam-Webster}, entry \textit{harness} (\url{https://www.merriam-webster.com/dictionary/harness}). The fighting-equipment sense is dated to around 1300 and the draught-animal sense to the early fourteenth century.} The verb \textit{to harness}, attested around 1300 in the sense of putting tack on an animal, only takes on the figurative sense that survives today, channeling a force to make it useful, in the late seventeenth century, as in \textit{harness the wind} or \textit{harness solar energy}.\footnote{The figurative sense of \textit{to harness}, controlling a force to extract useful work from it, is dated to the 1690s by the \textit{Online Etymology Dictionary} (entry \textit{harness}, verb sense) and recorded in the same sources above.} The idea never changes: take a brute force and channel it safely to produce work. That is the parent metaphor of everything that follows. Force without direction on one side; infrastructure that connects, controls, and measures on the other.

\begin{figure}[h!]
    \centering
    \includegraphics[width=0.92\linewidth]{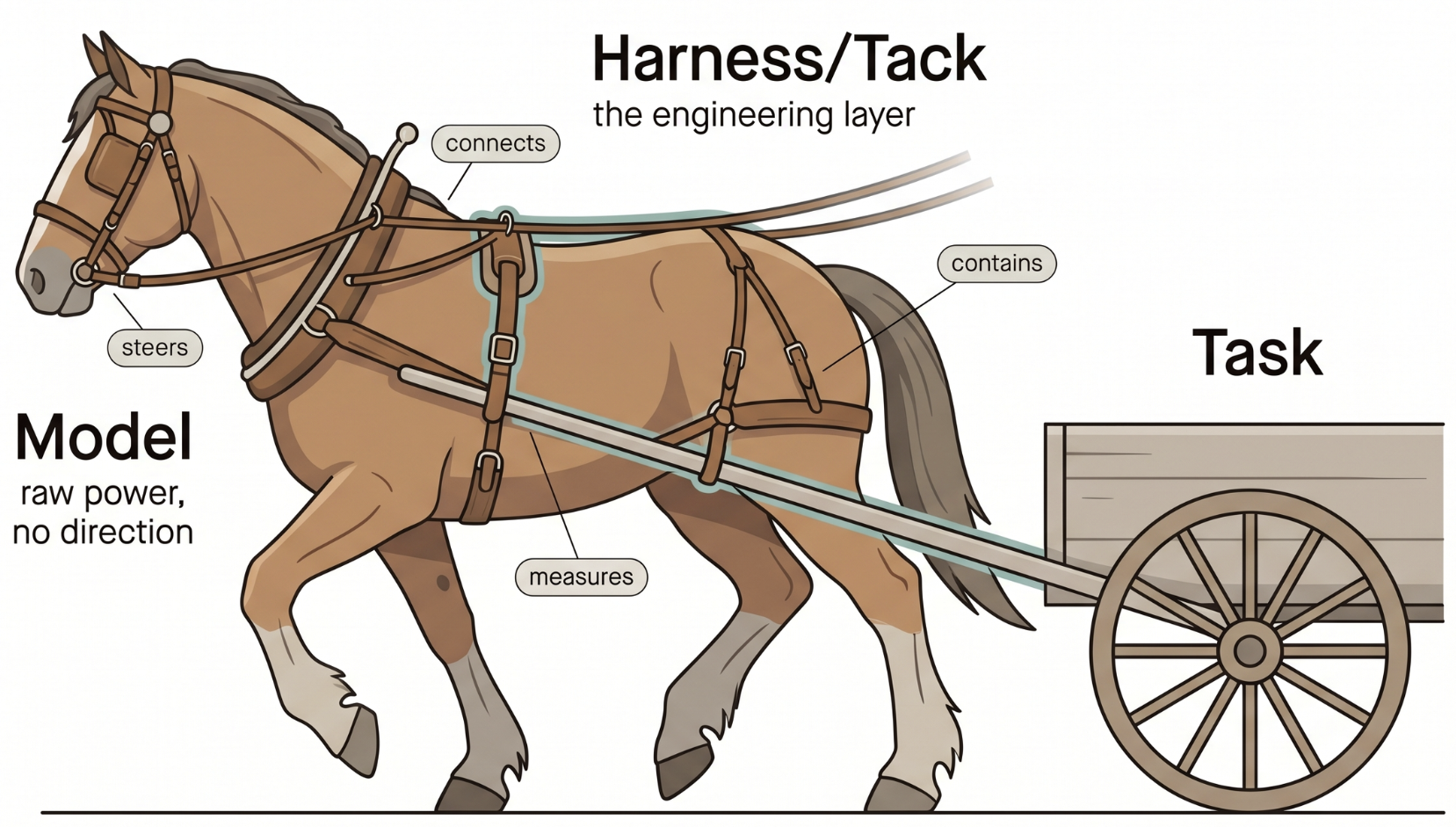}
    \caption{The parent metaphor of the harness. The language model is the horse, brute force with no direction of its own; the task is the cart to be pulled; the harness is the tack, the engineering layer that connects, steers, measures, and contains, turning uncontrolled power into trustworthy work.}
    \label{fig:arreio}
\end{figure}

Figure~\ref{fig:arreio} translates the metaphor to artificial intelligence. The horse is the model: power able to generate text and decide, but blind to the world and prone to bolting \cite{Bubeck2023Sparks}. The cart is the task to be delivered. The tack is the engineering layer that connects the model to the task, controls the execution, measures what happens, and monitors the result. The agent is the harnessed horse; the harness is the tack.

\subsection*{The classic test harness}
The term was not born with AI. In software testing, a \textit{test harness} is the set of scripts, mocks, stubs, and infrastructure that runs tests in a controlled and observable way. The tack metaphor was already there. Just as the tack connects and controls the horse, the test harness ties the code to an infrastructure that observes and controls its behavior during execution. This usage is well established in engineering, predates language models, and is the first reason the word sounds familiar to anyone who programs.\footnote{The classic sense of \textit{test harness} is described in consolidated software-testing vocabularies, such as the \textit{ISTQB Glossary} (\url{https://glossary.istqb.org}).}

\subsection*{The machine-learning evaluation harness}
From the test harness, close relatives derived in the world of models: the \textit{evaluation harness} and the \textit{benchmark harness}, for measuring performance. Here the harness is an evaluation suite. One connects the model to the harness, runs a set of standardized tasks, and measures how well it did. The focus is to evaluate, to assign a grade. That sense now dominates agent evaluation: SWE-bench calls its task executor a harness \cite{Jimenez2024SWEbench}, and benchmarks such as AgentBench, WebArena, Mind2Web, and \(\tau\)-bench follow the same logic, a scaffold that runs the system against scenarios and collects the result \cite{Liu2024AgentBench,Zhou2024WebArena,Deng2023Mind2Web,Yao2024TauBench}. It is a harness that observes from the outside, after the work has happened.

\subsection*{The agent harness}
The \textit{agent harness} inherits the name and the metaphor, and widens the scope. It does not just evaluate at the end. It controls, limits, verifies, and corrects the execution while it happens. The shift tracks a change of era. In classic machine learning the system was passive: input went in, output came out, and the harness measured quality. In the agentic regime the system acts, calls tools, browses, writes files, authenticates \cite{Yang2024SWEagent,Wang2024OpenHands}. A system that acts needs more than evaluation at the end. It needs control along the way. Table~\ref{tab:sentidos} summarizes the two senses that coexist today, whose confusion motivates this article.

\begin{table}[h!]
\centering
\small
\begin{tabularx}{\textwidth}{|l|Y|}
\hline
\textbf{Context} & \textbf{What \textit{harness} means} \\
\hline
Software testing (classic) & Scripts, mocks, and stubs that run code in a controlled and observable way \\
\hline
Model/agent evaluation & A suite that runs the system against standardized tasks and measures the result \\
\hline
Agent engineering (runtime) & The layer that controls, limits, verifies, and corrects the agent's execution while it happens \\
\hline
\end{tabularx}
\caption{The senses of \textit{harness} in circulation. The first two observe from the outside; the third, the target of this article, acts during execution.}
\label{tab:sentidos}
\end{table}

\begin{figure}[h!]
    \centering
    \includegraphics[width=0.98\linewidth]{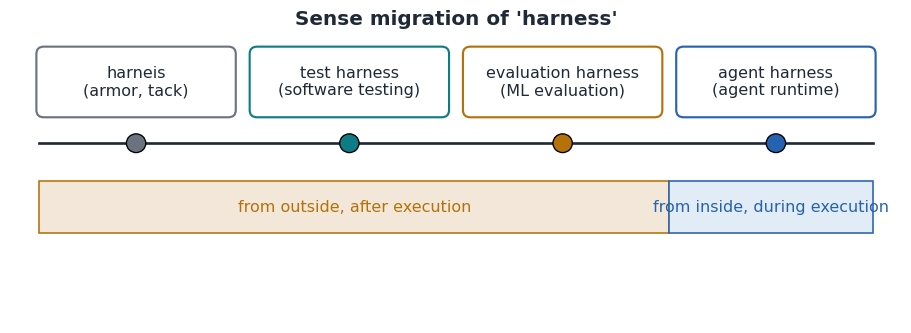}
    \caption{Timeline of the sense migration of \textit{harness}. From the Old French \textit{harneis} (armor and tack) to the software-engineering \textit{test harness}, to the machine-learning \textit{evaluation harness}, and to the agentic \textit{agent harness}. The lower band marks when control acts: from outside and afterward, in the first three; from inside and during, in the last.}
    \label{fig:timeline}
\end{figure}

Figure~\ref{fig:timeline} synthesizes this migration, from tack to agent, as a timeline. The thread is single: in every station, a harness is the infrastructure that channels a force (the horse, the code under test, the model) to produce useful work under control. What changes is what is channeled and when control acts. In the agent harness it acts at runtime, and that change is what demands a definition of its own, which we build next.

\section{Constitutive Definition (RQ2)}
\label{sec:definicao}

The genealogy tells us where the term comes from. What remains is to say what it designates today, and precisely. We define the agent harness by conditions, not by examples, and we turn that definition into a test that includes and excludes concrete cases: this is our answer to RQ2. A good constitutive definition enumerates what is necessary, drops what is incidental, and separates the concept from its neighbors \cite{Gruber1995Ontology}.

We propose the following reference definition of the agent harness.

\begin{quote}
\textit{An agent harness is the runtime engineering layer that wraps one or more language models and turns them into an agent able to accomplish tasks over an external environment, by coupling to the model: (i) an agent loop that interleaves reasoning, action, and observation; (ii) a tool interface that lets the model perceive and alter the environment; (iii) context management that decides what enters and leaves the model's window; and (iv) control mechanisms, that is, limits, verification, and deterministic actions, that make the execution more trustworthy, auditable, and contained.}
\end{quote}

A system is an agent harness if and only if it instantiates the four elements above at runtime. The definition states conditions, not cases. The temporal clause carries a lot of weight: the harness acts \textit{during} the task, and that is what separates it from the evaluation harness, which acts afterward. We did not invent the four elements; they come from the mechanisms the technical literature treats as recurrent in agents that act. The reasoning and action loop \cite{Yao2023ReAct,Shinn2023Reflexion}. Tool use and the interface with the environment \cite{Schick2023Toolformer,Yang2024SWEagent,Patil2023Gorilla}. Context management and retrieval \cite{Mei2025ContextEng,Liu2024LostMiddle,Lewis2020RAG}. And control by verification and containment \cite{Madaan2023SelfRefine,Rebedea2023NeMo}.

\subsection*{Anatomy}

\begin{figure}[h!]
    \centering
    \includegraphics[width=0.96\linewidth]{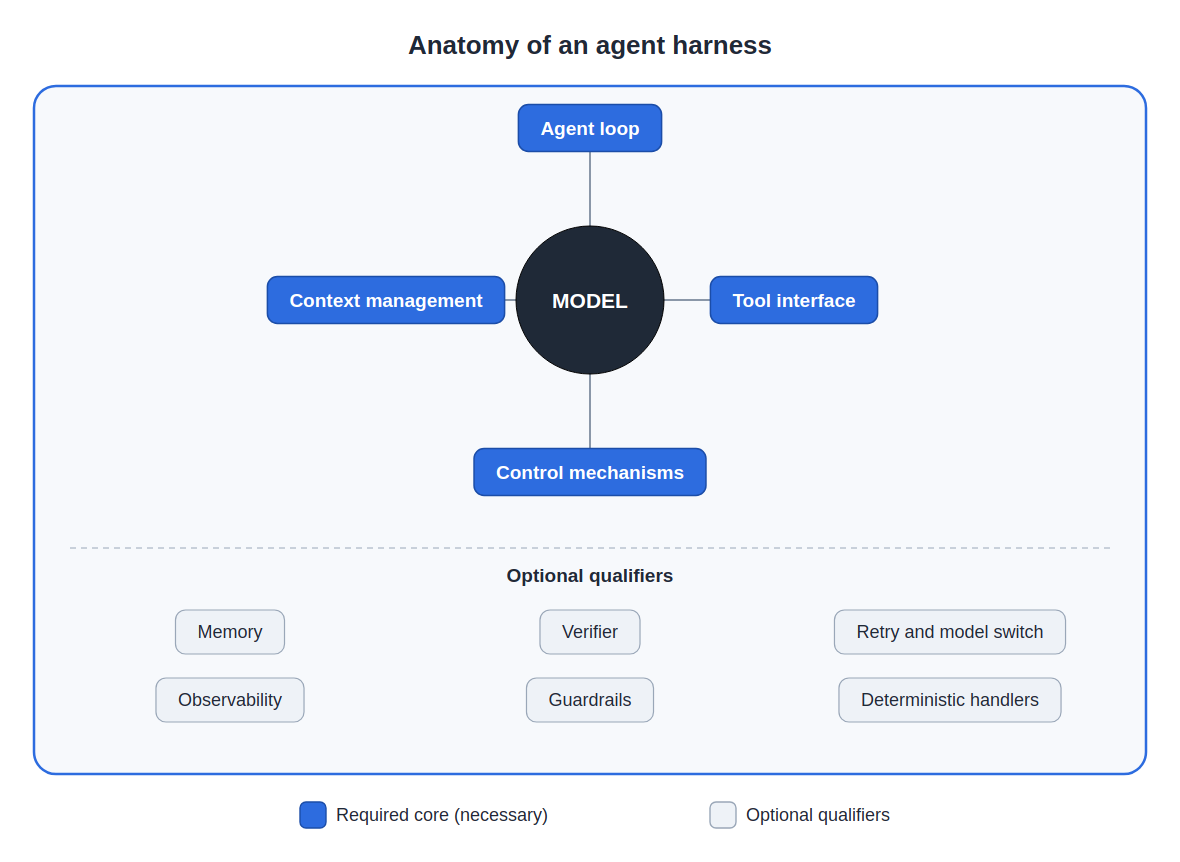}
    \caption{The anatomy of an agent harness. At the center, the model, the engine. Around it, the components that turn it into a trustworthy agent: tool registry, context manager, memory, agent loop, verifier, retry and model switch, observability, guardrails, and deterministic handlers. The four core elements (loop, tools, context, control) are necessary; the rest qualify the harness.}
    \label{fig:anatomia}
\end{figure}

Figure~\ref{fig:anatomia} opens the concept into components. At the center is the model, the engine. Around it, the harness assembles what the engine lacks to become a drivable vehicle: a \textit{tool registry} (the catalog of what the agent can do, read a file, run a command, browse), a \textit{context manager} (which compacts history and selects the relevant), a \textit{memory} (which survives across steps and sessions \cite{Packer2023MemGPT,Zhang2024MemorySurvey}), the \textit{agent loop} (reason, act, observe), a \textit{verifier} (which checks whether the task was actually accomplished, rather than accepting the model's word \cite{Madaan2023SelfRefine}), \textit{retry logic} with eventual model switching (which re-runs transient failures), \textit{observability} (an auditable record of what happened), \textit{guardrails} (safety limits \cite{Inan2023LlamaGuard,Rebedea2023NeMo}), and \textit{deterministic handlers} (ordinary code for the parts too sensitive for the model). Not every harness ships all of those optional components. The four elements of the definition, though, are the core. Without them there is no harness.

Each condition is necessary, and that becomes clear when we take one element away at a time and watch what is left. Without an agent loop, the system answers once and stops; it is a generator, not an agent, and no agent survey would classify it as such \cite{Wang2024SurveyAutonomous}. Without a tool interface, the model neither perceives nor alters the environment; it stays trapped in its own window, unable to act on a repository \cite{Yang2024SWEagent}. Without context management there is no viable loop: every multi-step task accumulates observations, tool outputs, and test results that compete for the model's window, and deciding what feeds back into the model is constitutive of the loop, not a luxury reserved for long tasks; the literature confirms that the alternative, letting useful information dilute in the buildup of history, degrades the model \cite{Liu2024LostMiddle,Mei2025ContextEng}. And without control mechanisms, the system acts, but no one can tell whether it did what it claims, nor is there a way to contain it; it is exactly the failure that motivates the topic, when an agent reports a nonexistent success and nothing verifies it \cite{Bubeck2023Sparks,Madaan2023SelfRefine}. The four, together, are sufficient. Any system that satisfies T1 through T4 already exhibits the behavior that motivated the concept: channeling the model's brute force with control exercised at runtime. Once the four hold, nothing essential is missing, and there is no fifth condition to add. Memory, verification, observability, and the like are not new elements; they are specializations of T1 through T4, that is, more robust ways to keep the loop, perceive the environment, manage context, and control execution, exactly as the next paragraph shows when it separates what is core from what is incidental. None of the four, in isolation, dispenses with the others.

What is NOT necessary also has to be said, and the definition stays lean because it leaves the incidental out. An agent harness does not require multi-agent: a single model in a loop with tools and control is already a harness, and multi-agent architectures are a design choice, not a requirement \cite{Qian2024ChatDev,Wu2023AutoGen}. It does not require learning or fine-tuning: the harness is engineering around the model, not modification of it. It does not require a specific model, and that is a property, not a defect; a good harness extracts trustworthy work from diverse models, and model switching is a control mechanism, not a dependency. Nor does it require a user interface: a harness can be a faceless library. Pulling these items into the definition would make it too narrow and would throw out systems that are, legitimately, harnesses.

\subsection*{The inclusion and exclusion test}

The definition becomes an instrument once we turn it into a decision procedure. Given a candidate system, ask, in order: (T1) does it maintain a loop that interleaves reasoning, action, and observation at runtime? (T2) does it offer the model an interface to perceive and alter an external environment? (T3) does it actively decide what enters and leaves the model's context? (T4) does it include at least one control mechanism, verification, limit, or deterministic action, that does not depend on the mere obedience of the model? A system is an agent harness if it answers \textit{yes} to T1 through T4. Fail any one and the system falls into a neighboring category; Section~\ref{sec:fronteira} shows which. Table~\ref{tab:teste} summarizes the test and anticipates what each failure implies.

\begin{table}[h!]
\centering
\small
\begin{tabularx}{\textwidth}{|c|Y|Y|}
\hline
\textbf{Test} & \textbf{Question} & \textbf{If the answer is \textit{no}} \\
\hline
T1 & Is there a reasoning, action, and observation loop at runtime? & Single-pass generator or fixed pipeline; not an agent \\
\hline
T2 & Is there a tool interface to perceive and alter the environment? & Isolated model or SDK that does not yet build the loop \\
\hline
T3 & Is there active management of what enters and leaves the context? & Naive wrapper that dumps history; brittle on long tasks \\
\hline
T4 & Is there at least one control mechanism independent of the model? & Demo without guarantee; trusts the model's word \\
\hline
\end{tabularx}
\caption{Inclusion and exclusion test derived from the constitutive definition. A system belongs to the concept if it answers \textit{yes} to T1 through T4; each failure indicates the neighboring category into which the system falls.}
\label{tab:teste}
\end{table}

Threshold and gradation are two caveats that keep the test from being read naively. The first is the threshold of each condition, especially T3. A wrapper that merely truncates history when it overflows the window does \textit{some} context handling, but it does not satisfy T3 in the intended sense. The verifiable criterion is this: T3 is satisfied if the decision about what enters and leaves the context depends on the \textit{content} of the task or of the current observation, and not merely on the \textit{size} of the buffer. A mechanical cut by size fails the test; active, task-aware selection that decides what matters for the current step passes. T2 also has a verifiable criterion: T2 is satisfied if the interface lets the model \textit{alter} an external environment (edit files, run commands), not merely read it or suggest text. Reading signals from the environment without being able to change it fails the test; an interface that acts on the environment passes. The same rigor holds for T4. The criterion is the following: a mechanism satisfies T4 if its effectiveness does \textit{not} depend on the model choosing to cooperate. A single log \textit{print} fails, because it does not alter the course of execution; a cap on tool calls or a deterministic check passes, because it contains or alters execution independently of whatever the model decides to do. The second caveat: membership is binary in its existence but gradual in its quality. A minimal loop that, at each step, re-runs the repository's test suite and declares success only if the suite passes satisfies T1 through T4 to the letter and is, in fact, an embryonic harness: re-running the tests is a genuine deterministic check that confirms the outcome regardless of whether the model claims to be done. What distinguishes it from Claude Code or OpenHands is not whether it belongs to the concept; it is the maturity of the mechanisms, above all the control ones. The test decides \textit{whether} a system is a harness. The anatomy qualifiers measure \textit{how} robust it is. Treating those two questions as one is the source of much of the terminological mess this article combats.

A harness and a guardrail must be contrasted, since the confusion between them is the most common of the topic. Guardrail and harness are not synonyms: the guardrail is a piece of the harness, not the whole. The difference is one of function. Guardrails \textit{limit}: they restrict, block, validate, impose boundaries, such as a cap on tool calls, a cost limit, or the blocking of destructive commands \cite{Rebedea2023NeMo,Inan2023LlamaGuard}. The harness, as a whole, \textit{enables}: the context manager helps the agent remember, memory avoids repeating work, retry overcomes transient failures, the verifier confirms completion. The test question is single: is this limiting the agent, or helping it execute? If it limits, it is a guardrail. If it helps, it is a harness. Guardrails are inside the harness; the harness is not inside the guardrails. For that reason the guardrail is not one of the lateral neighbors of the boundary treated in Section~\ref{sec:fronteira}, but a kind of control mechanism, part of T4: the relation here is part-whole, not that of two adjacent categories that get confused for each other. That asymmetry matters because it places control, not restriction, at the center of the concept.

\section{Boundary Delimitation (RQ3)}
\label{sec:fronteira}

A definition only discriminates if it separates the concept from its neighbors. Five concepts are routinely confused with the agent harness: agent framework, agent SDK, IDE plugin, eval harness, and orchestrator. We confront the harness with each one, always through the case that separates them, and we anchor that case in the inclusion and exclusion test of Section~\ref{sec:definicao}.

An agent framework differs from a harness by operating above it. An agent framework, such as AutoGen, CAMEL, AgentVerse, or the more widely adopted production frameworks CrewAI, LangGraph, Google's Agent Development Kit, and Agno,\footnote{Official sources for the production frameworks, cited here as grey literature: CrewAI (\url{https://crewai.com}); LangGraph (\url{https://www.langchain.com/langgraph}); Google's Agent Development Kit (\url{https://adk.dev}); Agno (\url{https://github.com/agno-agi/agno}).} offers abstractions to \textit{compose} agents: roles, conversation protocols, collaboration patterns \cite{Wu2023AutoGen,Li2023CAMEL,Chen2023AgentVerse,Hong2024MetaGPT}. The harness is the layer \textit{below}: it makes an individual agent act reliably on the environment. A framework can use a harness beneath each agent it coordinates. A harness needs no framework at all, since a single agent already instantiates it. The confusion arises because many frameworks embed a minimal harness. Two cases separate them: a system that merely routes messages between personas, without a loop of action on an external environment, fails T2 and is a framework, not a harness, whereas a single agent that edits files and verifies the result, without any role composition, is a harness, not a framework.

An agent SDK separates itself from a harness by being raw material, not a finished product. An SDK delivers the building blocks, function calling, tool definitions, loop primitives, but does not impose a running agent. It is raw material \cite{Patil2023Gorilla,Qin2024ToolLLM}. The harness is the product assembled from that material: it builds and runs the loop. A library that exposes the tool-calling primitive but leaves the developer to assemble the loop and the control fails T1, because there is no loop at runtime until someone writes it; it is an SDK. When that same library already delivers the closed loop, with context and verification, it has crossed the boundary and become a harness.

An IDE plugin is distinguished from a harness by closing no loop and not acting on the repository. An autocomplete plugin generates snippets from the cursor. It suggests code in the editor without maintaining task state, without acting on the repository, and without verifying the result \cite{Chen2021Codex}. It lacks the loop (T1) and the control (T4). The harness, even when it lives inside an editor, maintains the agent loop and acts on the environment. Completing the line where the cursor sits is a plugin; taking a task, planning, editing multiple files, running tests, and checking is a harness, even though both live in the same IDE.

The eval harness is the most direct name collision with the agent harness, and the easiest one to slip on. The eval harness, such as the one in SWE-bench, runs an agent against a set of tasks and \textit{measures} the result after each task has ended \cite{Jimenez2024SWEbench,Liu2024AgentBench,Yao2024TauBench}. One might object that it too has a loop, since it runs many tasks in sequence. The objection conflates two distinct loops. The eval harness's outer loop iterates over \textit{tasks} and collects scores. The loop of T1 is the inner loop of reasoning, action, and observation that conducts a \textit{single} task. The eval harness does not close that inner loop: it outsources it to the system under test, observes the outcome, and assigns a grade. From the point of view of a single task, it fails T1, because it does not decide the next step from the observation of the previous one, it only records what the system under test decided. The temporal clause seals the distinction: the eval harness acts \textit{after} the fact, the agent harness \textit{during} the fact. One judges the race, the other is the vehicle that runs it.

An orchestrator separates itself from a harness by following a fixed graph rather than an adaptive loop. An orchestrator, in the workflow sense, chains predefined steps in a fixed graph, a deterministic pipeline of steps \cite{Shen2023HuggingGPT}. The harness may contain orchestration, but what characterizes it is the adaptive loop: the next step depends on the observation of the previous one, not on a fixed graph. A pipeline that always runs A, then B, then C, without letting the observation alter the course, is an orchestrator; a loop in which the agent decides the next step from what it observed is a harness. Whatever only chains fixed steps fails T1, the adaptive loop.

As a boundary synthesis, Table~\ref{tab:fronteira} consolidates the five distinctions through the lens of the inclusion and exclusion test. The test is binary: each condition takes only \textit{yes} or \textit{no}, with no intermediate values, and a system passes a condition only if it satisfies it fully. Read vertically, it shows that no neighboring concept satisfies T1 through T4 at once in the sense of the agent harness. The boundary is not arbitrary: each neighbor fails, in binary fashion, at least one identifiable condition. That is what makes the definition an instrument, and not a label.

\begin{table}[h!]
\centering
\small
\begin{tabularx}{\textwidth}{|l|c|c|c|c|Y|}
\hline
\textbf{Concept} & \textbf{T1} & \textbf{T2} & \textbf{T3} & \textbf{T4} & \textbf{What separates it from the harness} \\
\hline
Agent harness & yes & yes & yes & yes & (it is the reference concept) \\
\hline
Agent framework & no & no$^{*}$ & no & no & Composes agents; the loop, tools, context, and control belong to the harness it coordinates, not to the framework itself \\
\hline
Agent SDK & no & yes & no & no & Offers blocks; assembles no loop at runtime \\
\hline
IDE plugin & no & no & no & no & Suggests code from the cursor without altering the repository or closing a loop \\
\hline
Eval harness & no & yes & no & no & Runs commands and applies patches, but measures from outside, after execution: fails T1 \\
\hline
Orchestrator & no & yes & no & no & Chains fixed steps; selection is not observation-driven and the loop is not adaptive (fails T1 and T3) \\
\hline
\end{tabularx}
\caption{Boundary of the agent harness against five neighboring concepts, through the lens of the T1 to T4 test. $^{*}$A framework may embed a harness in each agent it coordinates; the \textit{no} mark refers to the framework itself, not to the embedded harness.}
\label{tab:fronteira}
\end{table}

\section{Application to Real Harnesses (RQ4)}
\label{sec:aplicacao}

A definition that does not classify real cases is decorative. We apply the inclusion and exclusion test to six real harnesses and answer RQ4. The six were chosen to stress the definition across systems that differ from one another: Claude Code, Codex CLI, Aider, Cline, OpenHands, and SWE-agent. We classify each one from public documentation and, where available, from the academic description with a persistent identifier. We do not want to rank them. We want to show that the definition discriminates consistently across systems that differ widely from one another.

Claude Code is a terminal agent whose own documentation names the case a harness: it takes tasks, reads and edits files, runs commands, and maintains a reasoning and action loop, with context management and interception of dangerous operations at runtime.\footnote{Official Claude Code documentation: \url{https://code.claude.com/docs}. The \textit{Agentic harness} entry explicitly states that the product is the harness and the model is what is inside it.} It satisfies T1 (agent loop), T2 (file and shell tools), T3 (context compaction and selection), and T4 (runtime guardrails that ask for confirmation before destructive actions). Classification: agent harness, with all four core elements present.

Codex CLI is an open-source terminal agent that operates over the local repository with an action loop and a tool interface, offering permission modes that grade how much the agent can do without approval.\footnote{Codex CLI documentation and repository: \url{https://github.com/openai/codex}.} It satisfies T1 through T4, with control appearing explicitly in the permission modes (a form of runtime guardrail). Classification: agent harness.

Aider is a pair programmer in the terminal, integrated with version control. It edits files and commits changes with automatic commit messages, keeping a repository map as context management.\footnote{Aider documentation: \url{https://aider.chat}.} It satisfies T1 (an edit and correct loop), T2 (file editing and execution), T3 (a repository map that selects relevant context), and T4 (version-control integration, which makes each step auditable and reversible). Classification: agent harness, with auditability anchored in version control.

Cline is an agent that lives inside the editor (an IDE extension), with a plan and act loop, an interface to read and write files and run commands, and a flow that asks for human approval before sensitive actions.\footnote{Cline documentation and repository: \url{https://github.com/cline/cline}.} The case is instructive because it lives in an IDE, like a plugin, but it is not an autocomplete plugin: it maintains the loop (T1) and acts on the environment (T2), with managed context (T3) and human approval as control (T4). Classification: agent harness, and not an IDE plugin, precisely because it passes T1 and T4, where the autocomplete plugin fails.

OpenHands is an open platform for development agents, described academically. Its agents write code, use the command line, and browse the web, all inside an isolated environment (a sandbox) \cite{Wang2024OpenHands}. It satisfies T1 through T4, and sandboxed execution reinforces control as a strong form of deterministic containment. Classification: agent harness, with runtime containment as a distinguishing trait.

SWE-agent has as its central contribution the agent-computer interface: an interface designed so that the agent perceives and alters the software environment effectively, with an action loop on the repository \cite{Yang2024SWEagent}. It satisfies T1 (loop), T2 (the agent-computer interface is the materialization of T2), T3 (management of what the agent sees of the repository), and T4 (limits and an action format that structure the behavior). Classification: agent harness, and it is the case that best illustrates T2 in isolation. We should keep SWE-agent apart from SWE-bench, the eval harness against which systems such as it are measured \cite{Jimenez2024SWEbench}: the first executes, the second evaluates.

Classifying only systems already known to be harnesses confirms, it does not discriminate. The six above were chosen because they are harnesses, so passing the test was expected. The discriminating power appears when the test \textit{excludes}, and the most instructive case is a real, named system that is commonly grouped among AI coding tools. The first edge case is classic inline autocomplete, that is, code suggestion from the cursor, such as the completion feature of GitHub Copilot or Tabnine, scoped strictly to the inline completion feature.\footnote{Official product pages: GitHub Copilot \url{https://github.com/features/copilot} and Tabnine \url{https://www.tabnine.com}. The analysis here is restricted to the inline completion feature (suggestion from the cursor); it does not cover the agentic modes (\textit{agent mode}) these products have since come to offer, which could indeed satisfy the test and constitute a harness.} In this feature, the system suggests the continuation of the code without keeping task state. It reads editor signals but does not alter the repository, so it does not satisfy T2, and it does not close a reasoning and action loop over the repository (fails T1) and neither verifies the result nor contains the action (fails T4). The test excludes it, and rightly so: no one operates inline completion by handing it a multi-step task and expecting it to finish on its own. The second is a fixed orchestration pipeline, which always runs retrieve, then generate, then format, without letting the observation of one step alter the next. It has tools (T2), but its context selection is fixed rather than driven by the current observation (fails T3), and the course is a fixed graph, not an adaptive loop (fails T1). The test excludes it too. The two cases show that T1 through T4 is not a stamp that approves everything. It rejects plausible systems that lack the adaptive loop or the control, which is what separates a harness from an assistant or a pipeline.

\begin{figure}[h!]
    \centering
    \includegraphics[width=0.98\linewidth]{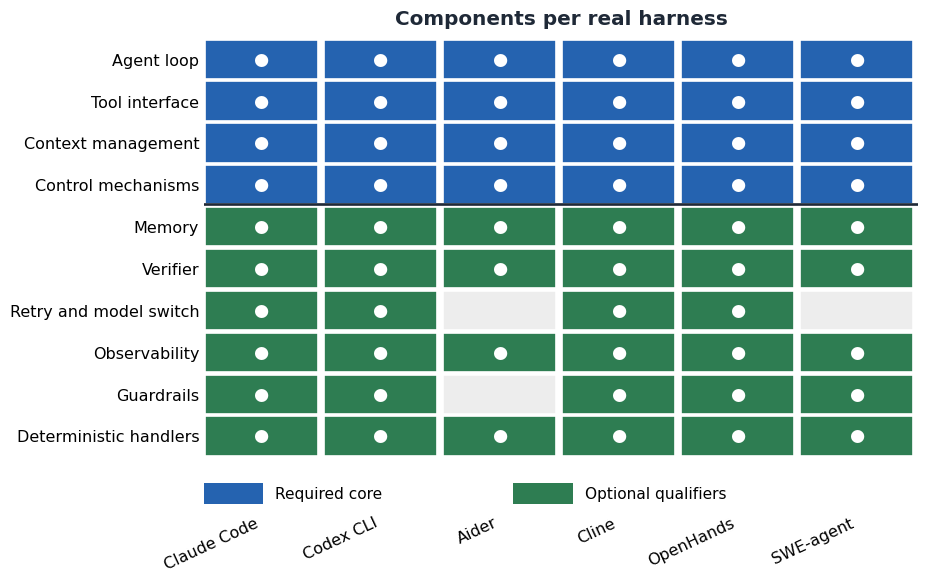}
    \caption{Component map per real harness. The rows are the anatomy components (necessary core above, optional qualifiers below); the columns are the six harnesses. The four core elements are present in all systems, as expected of a membership condition; the optional qualifiers vary and reveal design differences, above all in the form of control.}
    \label{fig:mapa}
\end{figure}

Figure~\ref{fig:mapa} maps the six systems against the anatomy components of Section~\ref{sec:definicao} and separates the necessary core from the optional qualifiers. Three things stand out. The four core elements (loop, tools, context, control) appear in all six, and that should not be over-interpreted: since the six were chosen for already being harnesses, the presence of the core is almost tautological, and the real discriminating power came from the edge cases the test excludes, not from the six passing. The optional components, by contrast, vary, and that is where the systems differentiate themselves in design: some emphasize sandbox containment (OpenHands), others auditability through version control (Aider), others runtime guardrails (Claude Code, Codex CLI). And the form of control (T4) is the axis of greatest variation, which anticipates the research agenda of Section~\ref{sec:agenda}. Table~\ref{tab:classificacao} closes the classification.

\begin{table}[h!]
\centering
\small
\begin{tabularx}{\textwidth}{|l|c|c|c|c|Y|}
\hline
\textbf{System} & \textbf{T1} & \textbf{T2} & \textbf{T3} & \textbf{T4} & \textbf{Distinguishing control trait (T4)} \\
\hline
Claude Code & yes & yes & yes & yes & Runtime guardrails; confirmation of destructive actions \\
\hline
Codex CLI & yes & yes & yes & yes & Graded permission modes \\
\hline
Aider & yes & yes & yes & yes & Auditability and reversal via version control \\
\hline
Cline & yes & yes & yes & yes & Human approval before sensitive actions \\
\hline
OpenHands & yes & yes & yes & yes & Sandboxed execution (deterministic containment) \\
\hline
SWE-agent & yes & yes & yes & yes & Structured action interface with limits \\
\hline
\end{tabularx}
\caption{Classification of the six real harnesses by the T1 to T4 test. All belong to the concept; the final column shows that the form of control (T4) is where the designs diverge most.}
\label{tab:classificacao}
\end{table}

\section{Design Axes and Research Agenda (RQ5)}
\label{sec:agenda}

\begin{figure}[h!]
    \centering
    \includegraphics[width=0.96\linewidth]{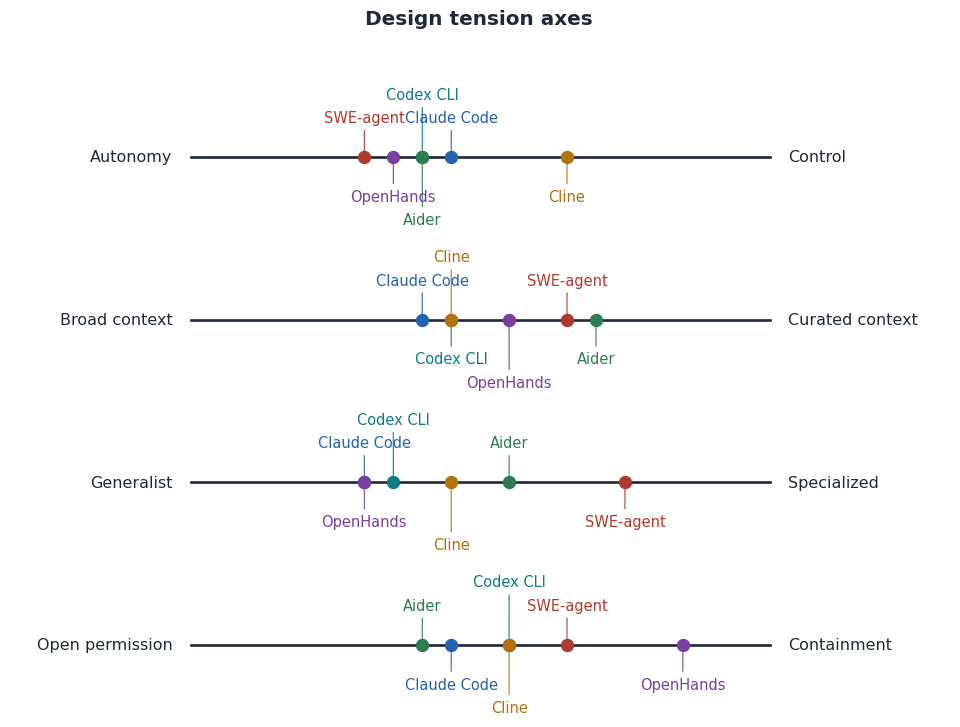}
    \caption{The six real harnesses placed qualitatively on four design tension axes: autonomy versus control, broad versus curated context, generalist versus specialized, and open permission versus containment. The positions derive from the classification in Section~\ref{sec:aplicacao} and illustrate that the concept is single while the design space is wide.}
    \label{fig:eixos}
\end{figure}

The six harnesses agree on the core and diverge on the qualifiers, as Section~\ref{sec:aplicacao} showed. These divergences are not noise. They are design choices about real tensions, and each tension opens a research front. We answer RQ5 by organizing the field along four tension axes and deriving, from each, open questions. Figure~\ref{fig:eixos} places the six systems on these axes qualitatively, based on the prior classification.

\subsection*{Axis 1: autonomy versus control}

The more the harness lets the agent decide on its own, the more useful and the more risky it becomes. Cline asks for human approval at every sensitive action and trades speed for safety; Codex CLI grades its permission modes and attempts a configurable middle ground. Agent failures, from prompt injection to adversarial behavior, show that autonomy without control independent of the model is brittle \cite{Liu2023PromptInjection,Debenedetti2024AgentDojo,Wei2023Jailbroken}. Two questions stay open. How do we measure the optimal point between autonomy and supervision for a task class? And how do we design verifiers that scale with autonomy, rather than bottlenecking it \cite{Madaan2023SelfRefine}?

\subsection*{Axis 2: broad versus curated context}

A harness can dump the whole repository into the context or actively curate what the model sees. Models degrade when useful information is diluted in long context, which favors curation \cite{Liu2024LostMiddle}; context engineering became a discipline precisely because of that \cite{Mei2025ContextEng}. But curating costs. It requires a repository map (Aider), well-calibrated retrieval (RAG) \cite{Lewis2020RAG}, and a memory that decides what to retain \cite{Packer2023MemGPT,Wang2024WorkflowMemory}. Two questions stay open. Which context-curation policies maximize performance per token? And how do we evaluate context management in isolation from the underlying model?

\subsection*{Axis 3: generalist versus specialized}

Generalist harnesses (OpenHands, Claude Code) seek to cover many tasks; specialized ones couple verifiers and handlers to the domain. The evidence suggests that effective control is problem-specific: the verifier that checks one business rule does not serve another. Two questions stay open. How much of the harness is reusable across domains and how much must be bespoke? And is there a generalist control core that composes with domain extensions?

\subsection*{Axis 4: open permission versus containment}

At the open extreme, the agent runs with the user's privileges; at the contained extreme, it operates in a sandbox (OpenHands) with a restricted surface. Deterministic containment is the strongest control because it does not depend on the model's cooperation, unlike the prompt guardrail, which is the weakest \cite{Barrett2023SecurityRisks,Inan2023LlamaGuard}. Two questions stay open. How do we compose strong containment with high utility, without the sandbox preventing the legitimate task? And which containment patterns transfer across harnesses?

Three findings cut across the four axes in a cross-cutting way. Control (T4) is what most distinguishes designs and what the formal literature has least consolidated; it is a particularly open front. The separation between model and harness carries a strategic consequence: the better the harness, the less the application depends on a single large and expensive model, since model switching becomes a control mechanism, not a rewrite \cite{Yehudai2025AgentEval}. If this separation between model and harness holds, it is plausible that the engineering differential may shift from the model toward the harness, since the harness is problem-specific; but this is a conjecture, and testing it empirically is left as future work. And the evaluation of harnesses, today, measures above all the model-harness pair through task benchmarks \cite{Jimenez2024SWEbench,Liu2024AgentBench,Yao2024TauBench}; an evaluation that isolates the harness's contribution, controlling for the model, is missing. That is, perhaps, a central methodological gap that this article's definition helps to make formulable: one can only measure a harness's contribution after knowing, precisely, what it is.

\section{Conclusion}
\label{sec:conclusao}

The term \textit{agent harness} became central to software engineering with generative AI before it became a definition. We reversed that order. We reconstructed the genealogy of the term, from the horse's tack to the test harness, to the evaluation harness, and to the agent harness. The thread is single: a harness is the infrastructure that channels a force to produce useful work under control. What makes the agentic sense distinct is one thing: this control acts at runtime. On that base, we define the harness by four conditions, agent loop, tool interface, context management, and control mechanisms, and we turn the definition into an inclusion and exclusion test that decides, given a concrete system, whether or not it is an agent harness.

The test is an instrument, not a label. At the boundary, it separates the harness from an agent framework, an SDK, an IDE plugin, an eval harness, and an orchestrator; each neighbor fails one condition you can point to. In practice, we classified six real harnesses, Claude Code, Codex CLI, Aider, Cline, OpenHands, and SWE-agent. All share the core. They diverge on the qualifiers, and they diverge above all in the form of control. From those divergences we drew a research agenda organized along four design tension axes.

At bottom, the contribution is one of conceptual hygiene: giving an overloaded term a meaning that includes and excludes cases. We wager that cumulative science about trustworthy agents depends, first of all, on a vocabulary that does not call too many things a \textit{harness}. The definition opens the question that motivated it and that it does not yet answer: how do we measure a harness's contribution by isolating it from the model it wraps? That is the natural continuation of this work. Knowing what the harness is was the necessary step toward knowing, next, how much it is worth.

\section*{Declaration on the Use of Generative AI}

The author conducted the research and wrote the manuscript. During the preparation of this study, however, the author used Grammarly tools to improve textual agreement and Claude Opus 4.8 to support text structuring and translation into English. After using these tools/services, the author reviewed and edited the content as needed and takes full responsibility for the content of the publication.

\bibliographystyle{plain}
\bibliography{refs}

\end{document}